\documentclass[preprint]{aastex}
\usepackage[english]{babel}
\usepackage{graphicx}
\usepackage[usenames,dvipsnames,svgnames,table]{xcolor}
\usepackage[colorlinks=true, citecolor=blue]{hyperref}
\usepackage{natbib}
\usepackage{gensymb}
\usepackage{savesym}
\savesymbol{tablenum}
\usepackage{siunitx}
\restoresymbol{SIX}{tablenum}
\usepackage[utf8]{inputenc}

\DeclareSIUnit\deg{deg}
\DeclareSIUnit\arcsec{arcsec}
\DeclareSIUnit\mas{mas}
\DeclareSIUnit\pixel{pixel}
\DeclareSIUnit\jansky{Jy}
\DeclareSIUnit\Jy{Jy}
\DeclareSIUnit\erg{ergs}
\DeclareSIUnit\ster{ster}
\DeclareSIUnit\beam{beam}
\DeclareSIUnit\arcmin{arcmin}
\DeclareSIUnit\parsec{pc}
\DeclareSIUnit\pc{pc}
\DeclareSIUnit\au{AU}
\DeclareSIUnit\lightyear{ly}
\DeclareSIUnit\year{yr}
\DeclareSIUnit\micron{\mu\meter}
\DeclareSIUnit\um{\mu\meter}
\DeclareSIUnit\Msun{\ensuremath{\textrm{M}_\odot}}
\DeclareSIUnit\Lsun{\ensuremath{\textrm{L}_\odot}}
\DeclareSIUnit\pm{\pico\meter}

\newcommand{\um}{$\mu$m}

\shorttitle{Haystacks: High-Fidelity Models of the Solar System}
\shortauthors{Roberge et al.}

\begin{document}

\title{Finding the Needles in the Haystacks: High-Fidelity Models of the Modern and Archean Solar System for Simulating Exoplanet Observations}

\author{Aki Roberge\altaffilmark{1},
Maxime J.~Rizzo\altaffilmark{1,9},
Andrew P.~Lincowski\altaffilmark{2,10},
Giada N. Arney\altaffilmark{3,10},
Christopher C.~Stark\altaffilmark{4},
Tyler D. Robinson\altaffilmark{5,10},
Gregory F. Snyder\altaffilmark{4},
Laurent Pueyo\altaffilmark{4},
Neil T.~Zimmerman\altaffilmark{1},
Tiffany Jansen\altaffilmark{6},
Erika R.~Nesvold\altaffilmark{7},
Victoria S.~Meadows\altaffilmark{2,10},
and
Margaret C.~Turnbull\altaffilmark{8}
}

\email{aki.roberge@nasa.gov, maxime.j.rizzo@nasa.gov, alinc@uw.edu, giada.n.arney@nasa.gov, cstark@stsci.edu, tydrobin@ucsc.edu, gsnyder@stsci.edu,
pueyo@stsci.edu, neil.zimmerman@nasa.gov, jansent@astro.columbia.edu, enesvold@carnegiescience.edu, vsm@astro.washington.edu, turnbull.maggie@gmail.com}

\altaffiltext{1}{Exoplanets and Stellar Astrophysics Lab, Code 667, NASA Goddard Space Flight Center, Greenbelt, MD 20771, USA}
\altaffiltext{2}{Dept.\ of Astronomy, University of Washington, Box 351580, Seattle, WA 98195, USA}
\altaffiltext{3} {Planetary Systems Lab, Code 693, NASA Goddard Space Flight Center, Greenbelt, MD 20771, USA}
\altaffiltext{4}{Space Telescope Science Institute, 3700 San Martin Drive, Baltimore, MD 21218, USA}
\altaffiltext{5}{Dept.\ of Astronomy and Astrophysics, University of California, Santa Cruz, CA 95064, USA}
\altaffiltext{6}{Dept.\ of Astronomy, Columbia University, Mail Code 5246
550 West 120th Street, New York, New York 10027}
\altaffiltext{7}{Dept.\ of Terrestrial Magnetism, Carnegie Institution for Science, 5241 Broad Branch Road, NW, Washington, DC 20015, USA}
\altaffiltext{8}{SETI Institute, 189 Bernardo Ave, Suite 200, Mountain View, CA 94043 USA}
\altaffiltext{9}{NASA Postdoctoral Fellow}
\altaffiltext{10}{NASA Astrobiology Institute Virtual Planetary Laboratory, University of Washington, Seattle, Washington, USA}


\begin{abstract}

We present two state-of-the-art models of the solar system, one corresponding to the present day and one to the Archean Eon 3.5~billion years ago. Each model contains spatial and spectral information for the star, the planets, and the interplanetary dust, extending to \SI{50}{\au} from the sun and covering the wavelength range \SI{0.3}{\micron} to \SI{2.5}{\micron}.
In addition, we created a spectral image cube representative of the astronomical backgrounds that will be seen behind deep observations of
extrasolar planetary systems, including galaxies and Milky Way stars. These models are intended as inputs to high-fidelity simulations of direct observations of exoplanetary systems using telescopes equipped with high-contrast capability. They will help improve the realism of observation and instrument parameters that are required inputs to statistical observatory yield calculations, as well as guide development of post-processing algorithms for telescopes capable of directly imaging Earth-like planets. 


\end{abstract}



\section{Introduction}

Do habitable planets and life exist in the nearby universe?  Scientists and the general public alike share intense interest in this question.  In recent years, the rapid pace of exoplanet discoveries has brought the exciting goal of finding habitable planets (or ``exoEarths'') and probing them for signs of life within reach.  Achieving this goal is exceedingly difficult and will likely require high-contrast direct imaging and spectroscopy of terrestrial planet atmospheres, at least for Earth-like planets around Sun-like stars \citep[e.g.][]{DesMarais:2002, Kaltenegger:2009}.  The great technical challenge is the need to tremendously suppress the light of the central star, while still allowing light from planets to be detected at small angular separations from the star.  Several designs for space telescopes with these capabilities are in development (e.g.\ LUVOIR\footnote{\url{http://asd.gsfc.nasa.gov/luvoir/}}, HabEx\footnote{\url{http://www.jpl.nasa.gov/habex/}}, and Exo-S\footnote{\url{http://exep.jpl.nasa.gov/stdt/}}).  However, these efforts are hampered by uncertainties and unknowns about the target systems, including the fraction of stars with terrestrial planets in their habitable zones ($\eta_\oplus$) and the characteristics of the interplanetary dust in the systems.  NASA's \emph{Kepler} mission has provided constraints on $\eta_\oplus$ \citep[e.g.][]{Burke:2015}, while a new survey with the Large Binocular Telescope Interferometer (LBTI) should provide the needed information on typical dust levels within the next several years \citep{Hinz:2016}.

Efforts to develop instrument and telescope designs are also hampered by overly simplistic simulations of the data to be acquired. Assessments of total mission science yields require statistical approaches founded on analytic formulae, due to the large number of astrophysical unknowns \citep[e.g.][]{Savransky:2010fd,Stark:2014dt}. However, those statistical calculations also demand a large number of observational input assumptions (e.g.\ required $S/N$ and spectral resolution values, contrast post-processing gains, and observational strategies). In most cases, these assumptions can and should be placed on a far more solid theoretical footing.

In particular, the impact of interplanetary dust within the target system, otherwise known as exozodiacal dust, on our ability to retrieve point source photometry has not been fully evaluated. 
A discussion of this problem and prospects for learning more about dust around nearby stars appears in \citet{Roberge:2012}. 
Several attempts have been made to assess exozodical dust impacts on exoplanet direct imaging observations, using approximate statistical methods \citep[e.g.][]{Savransky:2009, Stark:2014dt} or analytic prescriptions for the dust spatial structure \citep[e.g.][]{Noecker:2010}.
To date, no analysis has fully captured the realistic physical characteristics of exozodiacal dust (e.g.\ spatial structure, scattered light albedo) and attempted spectral extraction of planets in simulated high-contrast observations.

Furthermore, the problem of confusion between exoplanets and unresolved galactic and extragalactic background sources (i.e.\ stars and galaxies) has not been fully simulated and analyzed. 
The Earth has an apparent magnitude of $m_V \approx 30$ \citep[e.g.][]{Turnbull:2012} at \SI{10}{\pc}.
For comparison, the Hubble Ultra Deep Field (HUDF) has a limiting magnitude of $m_{AB} \sim 29$ for point-sources \citep{Beckwith:2006}.
So for high galactic latitude sight lines, an image deep enough to detect the Earth at \SI{10}{\pc} will be superimposed on a small portion of an HUDF-like field.  For other sight lines, interstellar material in the galactic plane will extinct background galaxies but a greater number of galactic background stars will be present.

To properly prepare for future exoEarth observations, we must better define the challenge, using the knowledge gained through decades of planetary and astronomical studies. Here we describe our work to create detailed models of our whole planetary system at two different epochs in its history: the modern solar system, which covers a period from \SI{2.5}{\giga\year} ago to the present day; and the Archean solar system, which covers a period between 3.5 and \SI{2.5}{\giga\year} ago. These two epochs are important because together they cover a large fraction of our solar system's estimated \SI{4.5}{\giga\year} lifetime, as well as almost the entirety of the period during which life has been present on the Earth. The modern solar system is a key benchmark since modern Earth is the linchpin of astronomers' definition of the stellar habitable zone \citep[e.g][]{Kopparapu:2013}. The Archean Earth represents our only other example of a planet where we know life was thriving. However, its atmosphere was devoid of the canonical oxygen biosignatures that are today's markers for the presence of life. Detecting the atmospheric features of anoxic worlds like the Archean Earth is thus an important exercise for future missions, as exoEarths could be observed at different times in their evolution.

Our high-fidelity models incorporate the current best knowledge of all major parts of the solar system: star, planets, and interplanetary dust. One unique aspect is that the 2D spatial models also carry spectral information for every component, allowing accurate simulations of multi-color imaging and planet spectroscopy. Another is that a modeled field of galactic and extragalactic sources, representative of the background scenes behind high galactic latitude target systems, has also been created.  This will enable the first rigorous tests of confusion problems in very high contrast direct exoplanet imaging.
Our spatial/spectral models are complementary to statistical exoplanet yield analyses, as they will help refine the observational parameters and strategies needed to achieve various science goals.

Section~\ref{sec:overview} briefly outlines the structure of the models, which are spectral image cubes. Details about the star and planet spectra appear in Section~\ref{sec:planets}. The 3D density models of interplanetary dust structures are discussed in Section~\ref{sec:dust}, as well as our assumptions about the dust properties required to turn those structures into surface brightness maps. Section~\ref{sec:back} explains how the spectral image cube containing the background sources was created. Finally, in Section~\ref{sec:sims}, we discuss use of the ``Haystacks'' modern solar system model to create a simulated observation with a high-contrast instrument concept. The solar system and background source spectral image cubes are publicly available\footnote{Download from \url{http://asd.gsfc.nasa.gov/projects/haystacks/haystacks.html} \label{foot:web}}; we encourage their use in other data simulations. 


\section{Overview of Model Content and Structure
\label{sec:overview}}

The solar system models consist of 3D datacubes, each slice representing the 2D flux density map at a different wavelength.
Each cube is assembled from several individual components computed separately: 1) a Sun spectrum in the central spatial pixel; 
2) planet spectra represented as single-pixel point sources at the appropriate locations (Section~\ref{sec:planets});
3) an inner solar system dust brightness distribution (Section~\ref{sub:innerdust}); 
4) an outer solar system dust brightness distribution (Section~\ref{sub:outerdust}).
The planets' locations are calculated from the planets’ orbital elements, the epoch of observation, and the viewing inclination (the epoch and inclination can be easily varied). 
The dust distributions are consistent with the planetary architecture (the planet masses and orbital elements) and the chosen viewing inclination.

The pixel values in all image slices are spectral flux densities in units of Jy.
A constant background from local zodiacal dust is added to each spectral image cube, corresponding to 23 mag/arcsec$^2$ in V-band (\SI{2.421e-6}{\Jy/\raiseto{2}\arcsec}). 
This represents an approximate mean value of the local zodiacal dust surface brightness for various viewing geometries \citep{Stark:2014dt}.
The wavelength dependence for the local zodiacal light is identical to that used for the extrasolar system inner dust cloud at \SI{1}{\au} from the star (see Section~\ref{sub:innerdust}).

The field-of-view (FOV) of each planetary system cube is $\SI{100}{\au} \times \SI{100}{\au}$ ($r_\mathrm{outer} = \SI{50}{\au}$), with $\SI{0.03}{\au} \times \SI{0.03}{\au}$ pixels. We generate the cubes assuming a distance to the system of 10~pc (\SI{10}{\arcsec} $\times$ \SI{10}{\arcsec} FOV; \SI{3}{\mas} $\times$ \SI{3}{\mas} pixels); the cubes can easily be scaled to other distances. 
We choose the image sampling such that there are $\sim 2$~pixels per spatial resolution element (Nyquist sampling) at \SI{0.27}{\micron} for a 9-m diameter telescope. 
For a 15-m diameter telescope, the cubes Nyquist sample the telescope point-spread function (PSF) at a wavelength of \SI{0.45}{\micron}.
At longer wavelengths, the cubes oversample the PSFs.
The spatial sampling of the cubes is sufficient for simulations of high-contrast observations with large aperture telescopes, such as the two LUVOIR observatories currently under study \citep[15-m and 9-m diameter telescopes;][]{Bolcar:2017gl}. Mission concepts with smaller telescopes \citep[such as HabEx;][]{Mennesson:2016fp} can easily bin the cubes down to their desired spatial resolution. 
Table~\ref{tab:sampling} shows the PSF sizes for various telescopes and the wavelength at which our cubes Nyquist sample the PSFs.

\begin{table}[ht]
\label{tab:sampling} 
\begin{tabular}{p{2.2in}|c|c|c|c|c|c}
\hline \hline
 & WFIRST & HabEx & HabEx & LUVOIR& HDST & LUVOIR\\
 & 2.4-m & 4-m & 6.5-m & 9-m & 12-m & 15-m \\
 \hline
 \hline
PSF ($\lambda/D$) at \SI{0.3}{\micron} (mas) & 25.8 & 15.5 & 9.5 & 6.9 & 5.2 & 4.1 \\
Solar system cubes, critical wavelength (\um) & 0.072 & 0.120 & 0.195 & 0.270 & 0.360 & 0.450 \\
Galaxy cubes, critical wavelength (\um) & 0.240 & 0.400 & 0.650 & 0.900 & 1.20 & 1.50 \\
\hline
\end{tabular}
\caption{Spatial sampling of telescope PSFs with Haystacks cubes. The critical wavelength is the wavelength at which the telescope PSF has two cube pixels per full-width at half maximum (Nyquist sampling).}

\end{table}

\begin{figure}[t!] \centering
\includegraphics[width=\textwidth]{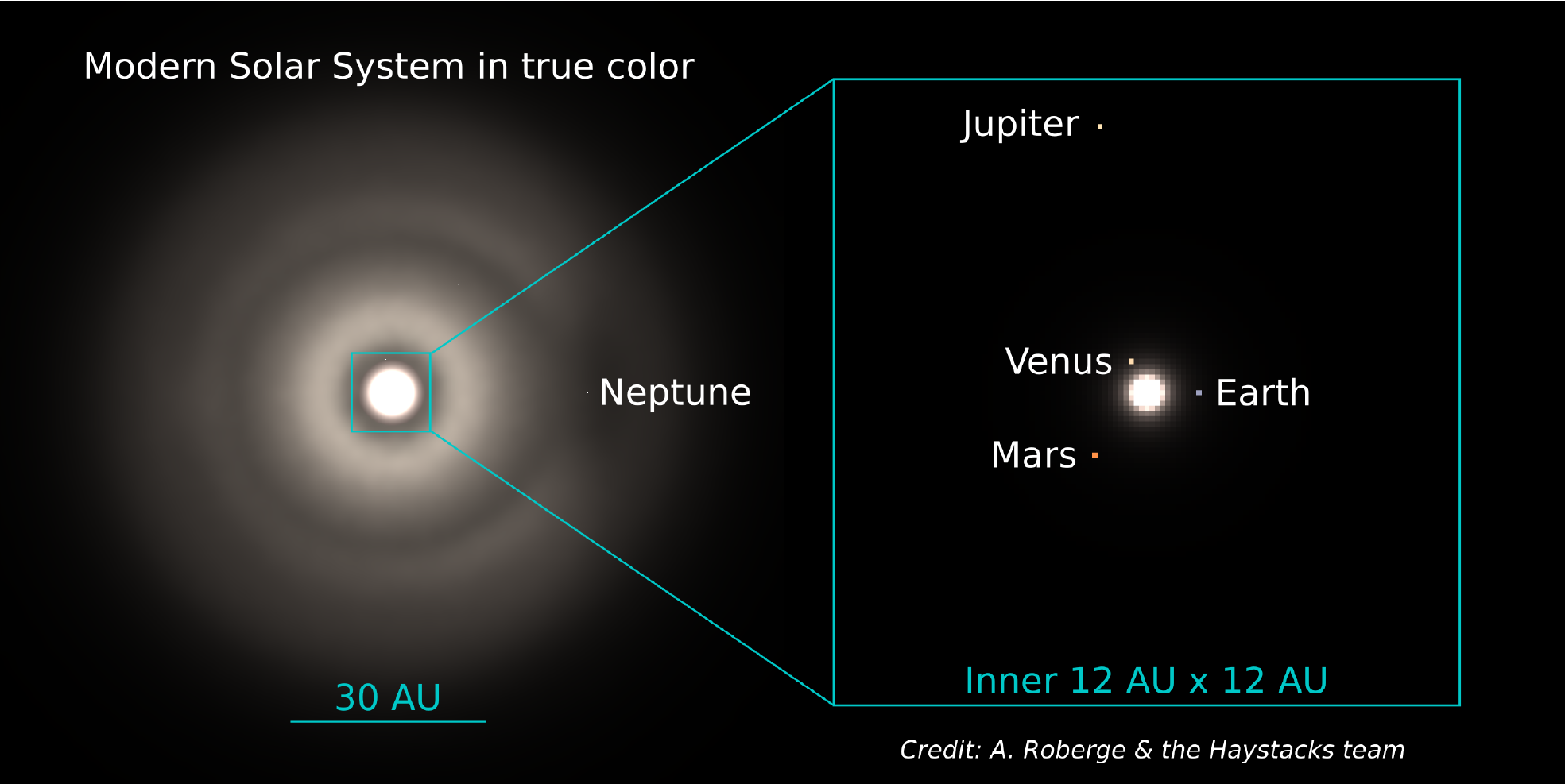}
\caption{True color image generated from our high-fidelity model of the face-on modern solar system. The left panel shows the entire field of view, while the inset panel zooms in on the inner system. For ease of viewing, the Sun and background sources were not included. The bright region at the center of the image is emission from exozodiacal dust. The planets occupy single pixels and are not readily apparent in the larger image without zooming in. The partial ring in the outer dust structure is caused by the dynamical influence of Neptune.
The colors were calculated using ColorPy (\url{http://markkness.net/colorpy/ColorPy.html}), which convolves the CIE~1931 empirically-based spectral response function of the human eye with the input spectrum in each pixel. Intensities were preserved in the full image, such that pixels can saturate in each color channel until turning white. In the inset panel, each planet was independently scaled to prevent saturation and preserve color. \label{fig:slices}}
\end{figure}

In the wavelength dimension, the cubes cover a broad bandpass from \SI{0.3}{\micron} to \SI{2.5}{\micron} at a spectral resolution $R=300$.
This resolution is about twice that needed for optimal detection of narrow atmospheric absorption features like the important O$_2$ A-band at \SI{0.76}{\micron} \citep[$R \approx 150$;][]{Brandt:2014gh}.
To keep the size of each cube file manageable, we divided the total spectral range into 10 smaller cubes with $\sim$20\% fractional bandwidth. The spectral resolution was set to $R=300$ at the central wavelength of each cube, and we adopted even wavelength spacing across the band.
Models for three viewing inclinations were constructed: 0$\degree$~(face-on), 60$\degree$, and 90$\degree$~(edge-on). Other system inclinations or epochs can be provided on request. A representative image slice from our modern solar system cube is shown in Figure~\ref{fig:slices}.

The spectral image cubes containing background galaxies have a FOV of \SI{12}{\arcsec} $\times$ \SI{36}{arcsec}, larger than that of the solar system cubes (\SI{10}{\arcsec} $\times$ \SI{10}{arcsec}, so that they may be superimposed on different parts of the galaxy field. 
The spatial resolution of the galaxy field is lower than that of the system cubes, for reasons explained in Section~\ref{sub:galaxies}. Therefore, the galaxy cubes have \SI{10}{\mas} $\times$ \SI{10}{\mas} pixels.
The background galaxy cubes are provided as separate files, so that they may be added to the planetary system cubes or not. We provide tools to spatially resample the size of the background pixels to match the pixel size of the system cubes if needed (note that this will not increase the actual spatial resolution of the galaxy field).

All cubes are provided in FITS format; each cube's header, as well as several additional FITS extensions, contain all of the information used to generate and combine the various cube elements. Each slice is contained in a different FITS extension to give the user maximum flexibility in computer memory management.
The codes to generate Haystacks cubes are publicly available\textsuperscript{\normalfont{\ref{foot:web}}}. However, computation of a consistent dust density structure given a multi-body planetary architecture requires separate, computationally-intensive codes \citep[e.g.][]{KuchnerStark:2010}.
Therefore, we encourage users who desire new Haystacks models with different planetary architectures and self-consistent dust distributions to contact the authors.

\section{Sun and Planet Spectra \label{sec:planets}} 

We initiated the process of building the solar system models using a Kurucz model for the solar spectrum\footnote{\url{ http://kurucz.harvard.edu/stars/sun/fsunallp.1000resam251}} (original spectral resolution $R = 1000$).  We used a combination of modeled and observed spectra to populate the planetary contribution to the spectral image cubes. The planets included are Venus, Earth, Mars, Jupiter, Saturn, Uranus, and Neptune. 
The planet flux densities were calculated from geometric albedo spectra using the appropriate phase angle given each planet's orbital position and assuming a Lambertian scattering phase function.

\subsection{Terrestrial Planets}

Geometric albedo spectra for the terrestrial planets (Venus, Earth, Mars) were created from reflectance models provided by the Virtual Planet Laboratory\footnote{\url{http://depts.washington.edu/naivpl/content/welcome-virtual-planetary-laboratory}}, generated using the Spectral Mapping Atmospheric Radiative Transfer (SMART) model \citep{Meadows:1996, Crisp:1997}.
The spectrum of the modern Earth is an equatorial view averaged over the 24-hour rotation period and verified with observations made during the EPOXI Earth flyby \citep{Robinson:2011}.
This is an acceptable generalization for other viewing geometries, as \citet{Cowan:2011} found that the spectral colors of Earth for a pole-on observation are similar to those of an equatorial view. 
The primary difference is a 20 - 30\% higher apparent albedo in the pole-on case due to the greater proportion of ice-covered surface. 
Thus, by using only the equatorial spectrum, the model represents a lower limit on modern Earth's apparent brightness in reflected light. 

The Archean Venus, Earth, and Mars spectra were also created using the SMART model. Whether early Venus was ever habitable is debated \Citep{Hamano:2013}. However, Venus's high present-day atmospheric D/H ratio suggests that it may have hosted a liquid water ocean when the Sun was fainter \citep{Donahue:1992}. The hypothetical early Venus spectrum we used assumed an Archean Earth-like atmospheric composition with biogenic gases removed and 100\% water cloud coverage; the bright albedo from clouds may be necessary to ensure low enough temperatures to avoid a runaway greenhouse effect on early Venus \Citep{Way:2016}. 
For Mars, we generated an Amazonian Period (starting 3 billion years ago) spectrum using an atmosphere model from \citet{Smith:2014}.  This version of Mars assumes an atmospheric composition similar to the modern planet but perturbed by low levels of volcanism. 

The Archean Earth atmosphere was self-consistently simulated using a 1-D coupled photochemical-climate model that includes microphysics for fractal hydrocarbon haze \citep{Arney:2016}. The photochemical portion is based on the model used in \citet{Zerkle:2012} to simulate hazes on early Earth, and the climate portion is based on the model used in \citet{Kopparapu:2013} to calculate habitable zones around different types of stars. We assumed a 1~bar atmosphere with 0.018~bars (50 times the present atmospheric level) of CO$_2$, in accordance with the CO$_2$ paleosol constraints of \citet{Driese:2011}. We adopted a CH$_4$/CO$_2$ ratio of 0.2, which is high enough to cause the formation of an organic haze, which laboratory, geochemical, and theoretical studies suggest existed during parts of the Archean \citep[e.g.][]{Trainer:2006, Izon:2015, DomagalGoldman:2008}. The Archean spectrum also included water clouds constructed as a weighted average of 50\% haze only, 25\% cirrus clouds and haze, and 25\% stratocumulus clouds and haze \citep{Robinson:2011}. The inclusion of organic haze on Archean Earth allows analysis of an inhabited planet radically different from modern Earth. 

\subsection{Giant Planets}

The Archean Eon is about 1~Gyr after the epoch of planet formation, when the thermal evolution of the giant planets was largely complete \citep{Fortney:2010}. Therefore, we used spectra of modern Jupiter, Saturn, Uranus, and Neptune for both epochs.
We decided to use observations instead of models, since the various giant planet model spectra we examined did a poor job fitting the depths of atmospheric absorption lines in observed spectra, likely due to inaccurate cloud or haze prescriptions.
For all the giant planets, we used observed albedo spectra from \citet{Karkoschka:1998} to cover the \SIrange{0.3}{0.9}{\micron} wavelength range. 
We adopted the full-disk albedo for Jupiter at $6.8\degree$ phase angle and Saturn at $5.7\degree$ phase angle as the geometric albedo (full-disk albedo at $0\degree$ phase angle) for these planets.
For Uranus and Neptune, \citet{Karkoschka:1998} provided geometric albedo spectra.

\begin{figure}[ht] \centering
\includegraphics[width=5in]{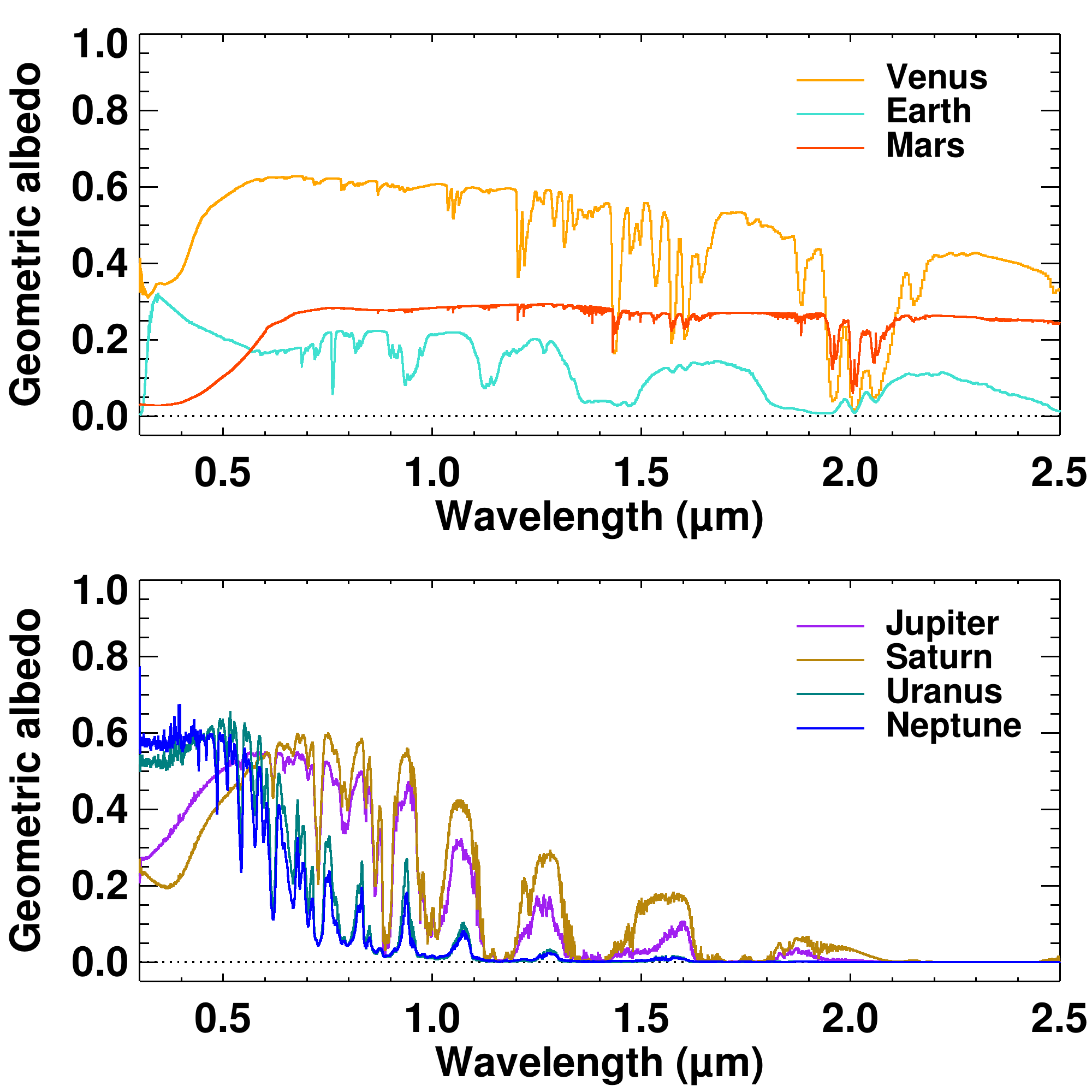}
\caption{Geometric albedo spectra of the seven planets included in the Haystacks modern solar system model. The spectra of Venus, Earth, and Mars are from models, while the spectra of Jupiter, Saturn, Uranus, and Neptune are from ground-based observations.}
\label{fig:planets1}
\end{figure}

\begin{figure}[ht] \centering
\includegraphics[width=5in]{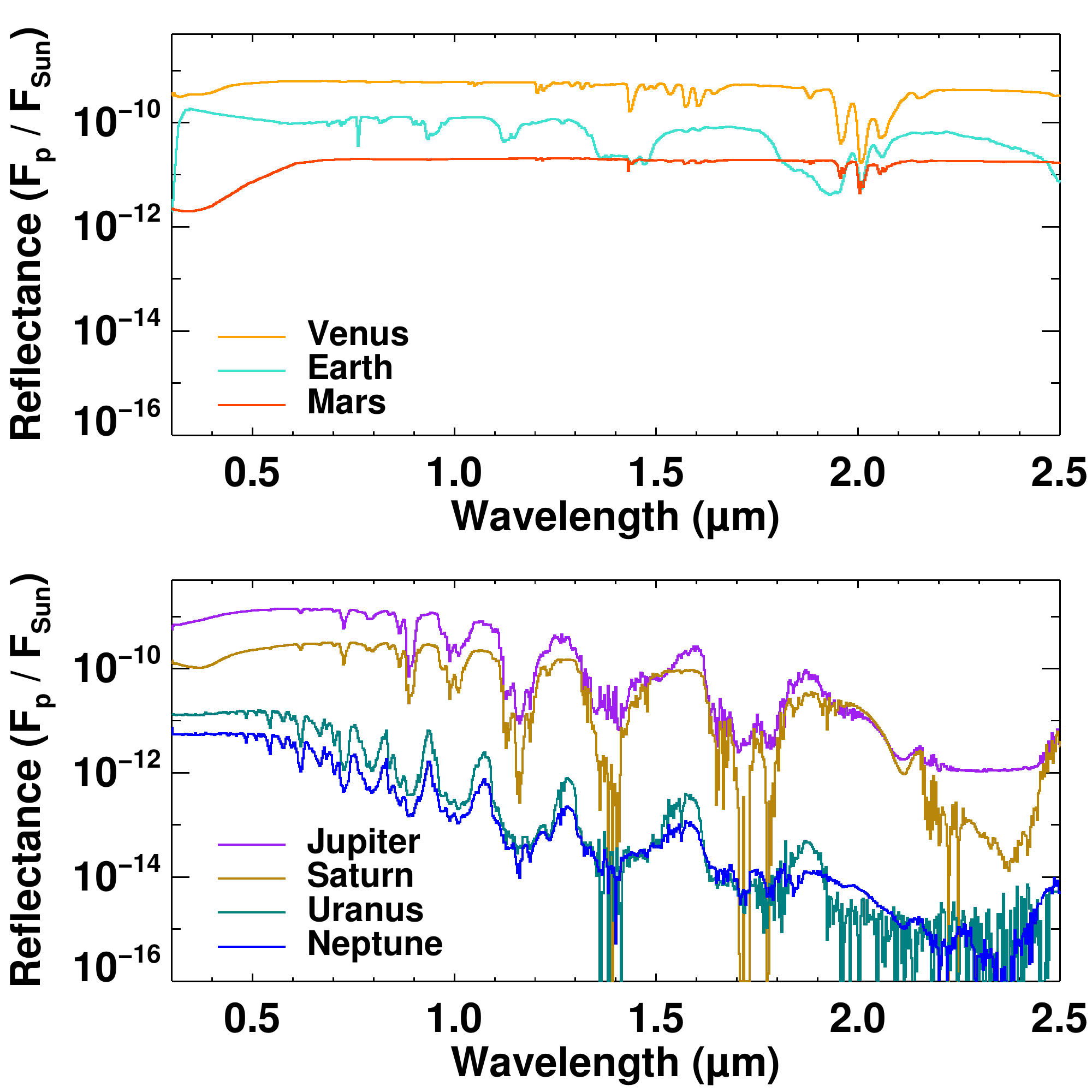}
\caption{Reflectance spectra of the seven planets included in the Haystacks modern solar system model (planet spectral flux density/Sun spectral flux density). All spectra plotted here were calculated for a phase angle of $90\degree$ (i.e.~quadrature).}
\label{fig:planets2}
\end{figure}

\begin{figure}[h!] \centering
\includegraphics[width=5in]{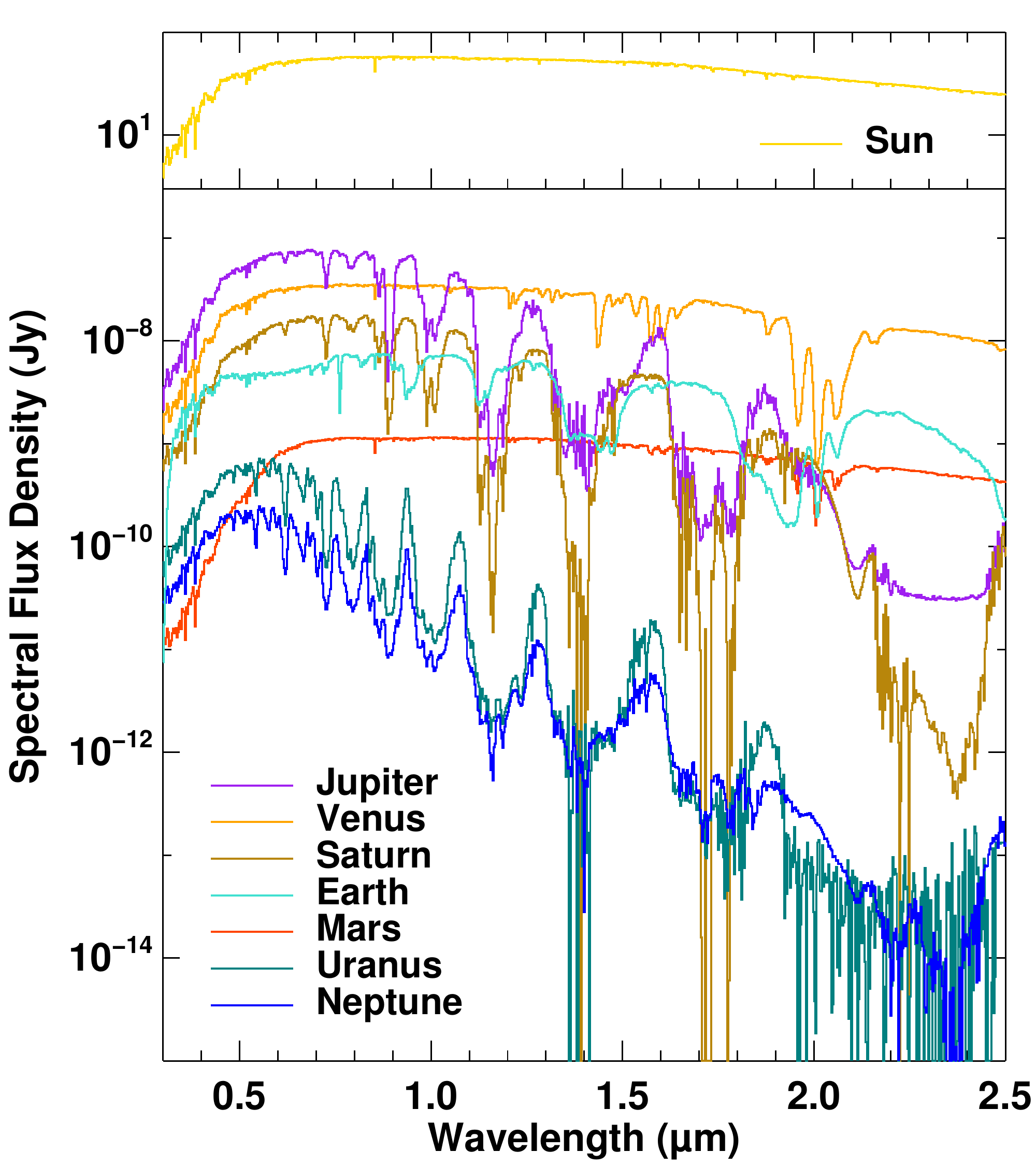}
\caption{Absolute spectral flux densities of the seven planets included in the Haystacks modern solar system model, along with the Sun spectrum in the upper panel. 
The solar system was placed at a distance of \SI{10}{\pc}. The planet spectra plotted here were calculated for a phase angle of \ang{90;;} (i.e.~quadrature).}
\label{fig:planets3}
\end{figure}

\begin{figure}[ht] \centering
\includegraphics[width=5in]{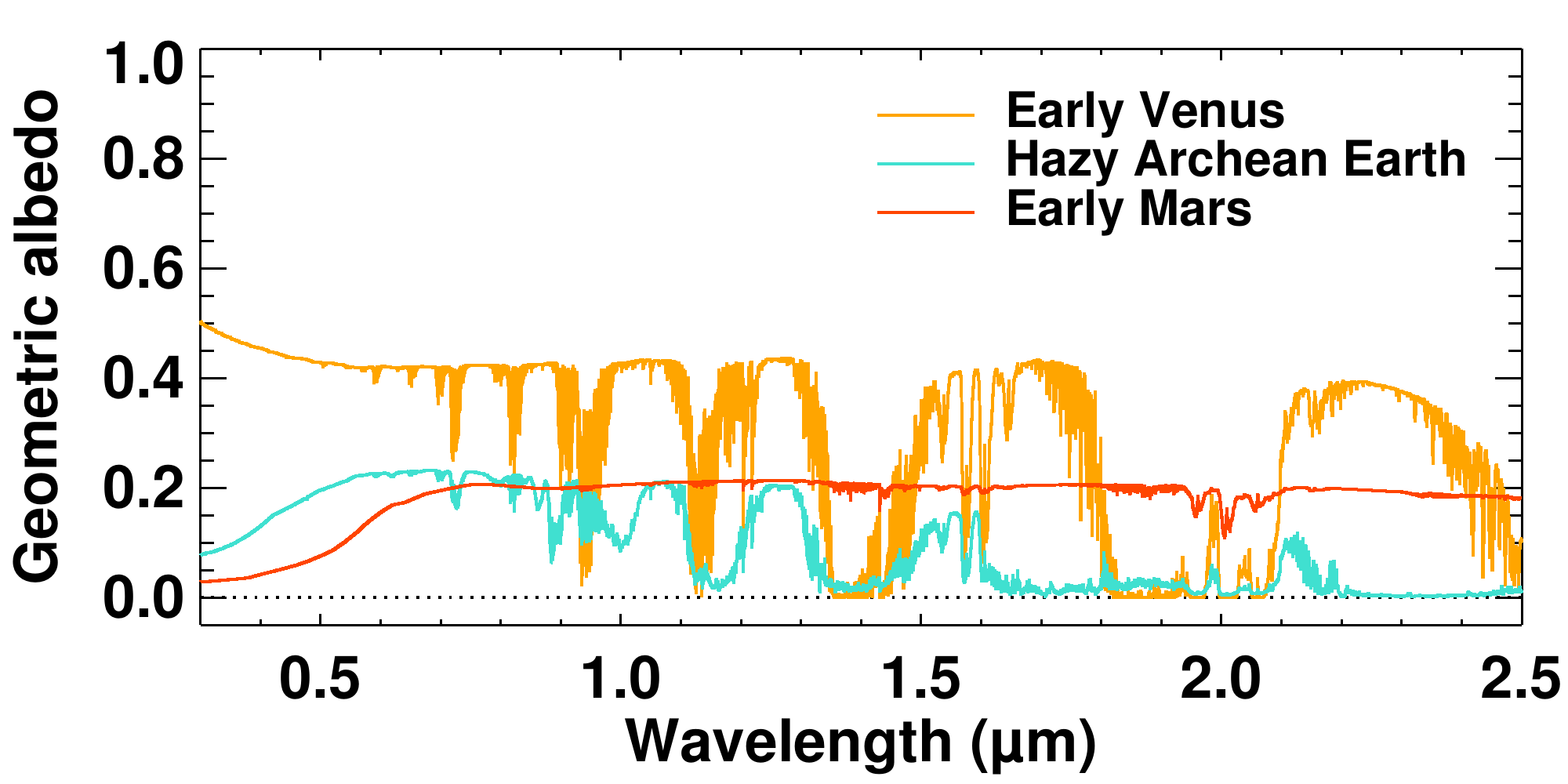}
\caption{Modeled geometric albedo spectra of the three terrestrial planets included in the Haystacks Archean solar system model.}
\label{fig:planets4}
\end{figure}

To cover the \SIrange{0.9}{2.5}{\micron} range for all the giant planets, we used observations obtained with the SpeX instrument on the NASA Infrared Telescope Facility\footnote{\url{http://irtfweb.ifa.hawaii.edu/~spex/IRTF_Spectral_Library/index_files/Planets.html}} \citep{Rayner:2009}.
The observed flux densities were divided by the Sun spectrum to create reflectance spectra, then scaled to match the \citet{Karkoschka:1998} albedo spectra in the wavelength region where the two datasets overlapped, resulting in approximate geometric albedo spectra. A \SI{0.06}{\micron} gap in wavelength coverage between the H and K bands (near \SI{1.85}{\micron}) in the IRTF spectra was patched with model giant planet spectra from \citet{Burrows:2004}; the detailed spectral shape in this region should be treated with caution.
Various spectra for the seven modern solar system planets are shown in Figures~\ref{fig:planets1}, \ref{fig:planets2}, and \ref{fig:planets3}. The geometric albedo spectra for the inner planets of Archean solar system are shown in Figure~\ref{fig:planets4}. 

\section{Interplanetary Dust Structure and Spectrum \label{sec:dust}}

In this section, we describe construction and assembly of the exozodiacal dust scattered light distributions. 
The dust component affects high-contrast observations of planets by adding photon noise, while the dust spatial structure can affect identification and extraction of point sources.
While both effects make it more difficult to detect point source planets, 
the dust spatial structure itself can indicate the presence of planets too faint to be directly detected \citep[e.g.][]{Kuchner:2003ja}.
The two models that we present have amounts of exozodiacal dust equal to that of the modern solar system and that expected for the Archean solar system. Should other values be of interest, we find that scaling the dust brightnesses by a factor of $\sim 10$ does not significantly impact the spatial distribution of the emission (C.~Stark, personal communication). For higher exozodiacal dust levels, we are able to construct new density maps upon request.

\subsection{Inner Dust Cloud: Zodiacal Dust \label{sub:innerdust}}

We simulated emission from the inner interplanetary dust using the ZODIPIC tool \citep{Kuchner:2012}, which uses the \citet{Kelsall:1998} model to make images of the solar zodiacal cloud. \citet{Kelsall:1998} fit an empirical model to detailed maps of the infrared sky at 10 wavelengths from \SI{1.25}{\micron} to \SI{240}{\micron} from the Diffuse Infrared Background Experiment (DIRBE) aboard the Cosmic Background Explorer satellite. 
They found four main components in the zodiacal cloud structure: the smooth dust cloud, the dust bands, Earth's resonant dust ring, and the Earth-trailing dust clump.
ZODIPIC evaluates the \citet{Kelsall:1998} model (hereafter the K98 model) and calculates the flux density contributions from both scattered light and thermal emission.

The scattered and emitted flux densities produced by ZODIPIC differ slightly from the those of the K98 model. \citet{Kelsall:1998} calculated the dust grain emissivity, assumed to be uniform throughout the zodiacal cloud, at 8 discrete wavelengths from \SI{3.5}{\micron} to \SI{240}{\micron}. ZODIPIC, which is designed to calculate the zodiacal cloud flux at any wavelength, uses an analytic fit to these emissivities. Additionally, \citet{Kelsall:1998} determined the albedo for each component of the zodiacal cloud separately, while ZODIPIC uses only the albedo of the smooth component and applies it to the dust bands, ring, and clump. These minor differences do not significantly impact the total zodiacal cloud flux, but do underestimate the flux of the bands and the ring by a factor of $\sim 2$ at $\lambda \gtrsim \SI{25}{\micron}$ (outside the bandpass of the Haystacks models). Additionally, ZODIPIC uses a wavelength-independent albedo chosen so that the polar sky brightness at V band is $\sim$23~mag$/$arcsec$^2$ at an ecliptic latitude of \ang{15;;} \citep{Cox:2000}.

Because the DIRBE data did not include wavelengths shorter than \SI{1.25}{\micron}, ZODIPIC further modifies the K98 model in the scattered light regime. It combines the 3D dust grain cross sectional density function from the K98 model with an analytic scattering phase function (SPF) to calculate the scattered light. For all our models, we choose to approximate the dust disk's SPF as a single Henyey Greenstein function with an asymmetry parameter appropriate for moderately forward-scattering grains ($g = 0.17$), as measured for Saturn's rings over a modest range of viewing angles \citep{Hedman:2015ie}.
This asymmetry parameter is within the range of values measured for debris disks  \citep[e.g][]{Schneider:2009fk}. As mentioned in \citet{Hedman:2015ie}, more complex SPFs are preferred to accurately estimate surface brightnesses for dust disks with inclinations close to edge-on. Our code can be adapted to model different SPFs upon request.

Most of the flux from the inner dust cloud comes from the central region, which has a size scale orders of magnitude smaller than the cloud's outer radius; therefore, ZODIPIC uses a simple mesh refinement scheme with logarithmic radial intervals. This allows ZODIPIC to synthesize images of the structure of the inner dust cloud while correctly calculating the total flux to an accuracy of $\sim 1\%$ or better. The inner radius of the dust model is set to the dust sublimation radius, assuming a dust sublimation temperature of \SI{1500}{\kelvin}.
ZODIPIC calculates the sublimation radius using the K98 model temperature law, $T(R) = T_0 \: (R/1 \ {\rm AU})^{-\delta}$. With $T_0 = \SI{286}{\kelvin}$ and $\delta=0.467$ for the Sun, a dust sublimation temperature of \SI{1500}{\kelvin} corresponds to an inner radius of \SI{0.03}{\au}.

The outer radius of the inner dust structure was set to \SI{10}{\au}. The K98 model is only valid out to \SI{3.28}{\au} (the location of Jupiter's 2:1 mean motion resonance), due to a lack of data to constrain the dust structure at larger distances. In the face of our current uncertainty about the solar system dust between \SI{3.28}{\au} and \SI{10}{\au} (the inner edge of the outer dust model described below), we choose to simply extrapolate the K98 model to cover this region.
We set ZODIPIC to include the K98 models of Earth's resonant ring and trailing dust clump, as well as three asteroidal dust bands. The face-on optical depth of the cloud is set to the K98 model value of $7.11 \times 10^{-8} (r/1 \ \mathrm{AU})^{-0.34}$.

\subsection{Outer Dust Cloud: Kuiper Belt Dust \label{sub:outerdust}}

Observations of the Kuiper Belt dust cloud are limited to estimates of the overall density from dust counting instruments on board missions like \emph{Pioneer 10}, \emph{Pioneer 11}, and \emph{New Horizons}.  We must therefore rely on dynamical models of the Kuiper Belt dust cloud to determine its brightness and structure. We used the 3D models of \citet{KuchnerStark:2010}, hereafter referred to as KS10, which self-consistently include the effects of resonant dynamical perturbations by planets and grain-grain collisions, both of which have a significant impact on the morphology of the Kuiper Belt dust cloud. A detailed description of these models appears in KS10; here we provide a brief summary.

KS10 modeled the Kuiper Belt dust cloud by numerically integrating the equations of motion for \num{75000} test particles subject to gravity from the Sun, Jupiter, Saturn, Uranus, and Neptune, plus the forces of radiation pressure, Poynting-Robertson drag, and solar wind drag. KS10 assumed the dust originated from three populations of Kuiper Belt Objects (KBOs): dynamically hot KBOs, dynamically cold KBOs, and plutinos. The dust production rate of each KBO population was set proportional to that population's estimated mass, such that hot KBOs produced 79.7\% of all dust, cold KBOs produced 16.3\%, and plutinos produced 4\%.  

In the KS10 modeling, particles migrated inward due to drag forces after ejection from the source bodies.  Neptune trapped many of these particles into exterior mean motion resonances (MMRs), creating a clumpy resonant ring structure that rotates with the planet.  Saturn also trapped particles into exterior MMRs, but because KS10 recorded the particle positions in the frame co-rotating with Neptune, Saturn's resonant ring structure is azimuthally smoothed and only Neptune's resonant ring structure is resolved in these models.  Particles were removed from the integration once they reached a semi-major axis $a < \SI{2.5}{\au}$, but the vast majority of particles were ejected by Jupiter and Saturn prior to reaching this distance, creating a natural inner edge to the Kuiper belt dust cloud at $\sim\SI{10}{\au}$.  Particles with $a > \SI{300}{\au}$ were also removed; the models of KS10 are only valid out to a circumstellar distance of $\sim\SI{100}{\au}$.

The KS10 modeling included 25 values for the ratio of the force due to radiation pressure to the force due to the Sun's gravity, ranging from $\beta \sim 0.0005$ to $\beta \sim0.5$. Assuming spherical icy grains with a radiation pressure coefficient $Q_{\rm PR} = 1$ and a grain density $\rho=\SI{1}{\gram\per\raiseto{3}\centi\meter}$, the range of $\beta$ values corresponds to grain sizes ranging from $\sim1$ to $\sim\SI{1200}{\micron}$.  KS10 assumed that the size distribution of dust at the time of production followed a Dohnanyi size distribution, which is appropriate for bodies in a collisional cascade \citep{Dohnanyi:1969}.  KS10 then applied the collisional grooming algorithm in \citet{StarkKuchner:2009} to account for the collisional destruction of grains via grain-grain collisions.  The collisional grooming algorithm self-consistently calculated the size distribution at all points within the disk, producing a final grain size distribution that differed from a Dohnanyi distribution.

Using the collisional grooming algorithm, KS10 produced 4 models of the Kuiper Belt dust cloud: one with a maximum geometric optical depth $\sim10^{-7}$ (consistent with the observations of \emph{Pioneer 10} and \emph{11}) and 3 denser models with maximum optical depths $\sim10^{-6}$, $\sim10^{-5}$, and $\sim10^{-4}$.  As shown by Figure~8 in KS10, models with optical depths $\le 10^{-6}$ feature a prominant clumpy resonant ring structure created by Neptune and a minor resonant ring created by Saturn (though it is not resolved by the model). Grain-grain collisions smooth these structures, such that models with optical depths $\ge 10^{-5}$ take on the appearance of an azimuthally symmetric ring of dust at the location of the parent bodies.

To synthesize images of the KS10 Kuiper Belt dust cloud density models, we used the publicly available \emph{dustmap}\footnote{Download from  \url{http://www.starkspace.com/code/}} radiative transfer package \citep{Stark:2011}. The \emph{dustmap} code takes a set of 3D positions of dust particles, illuminates them with starlight, and synthesizes an image including scattered light and thermal emission. The user defines the distance and orientation of the system, the stellar parameters, imaging parameters (spatial resolution, FOV, wavelengths), dust grain sizes, and the scattering phase function. As for the inner solar system dust, we assume that the latter is a Henyey Greenstein scattering phase function with an asymmetry parameter $g = 0.17$ (see Section~\ref{sub:innerdust}). 
The user must also provide \emph{dustmap} with optical constants for the dust as a function of size, or identify the grain composition to adopt optical constants calculated within the code using Mie Theory.
We used the latter option and assumed grains composed of pure water ice. Such grains have relatively high albedos compared to grains of other compositions, and the surface brightness of our Kuiper Belt dust represents an upper limit.  

\subsection{Scaling the Dust for the Archean Solar System}

\emph{Spitzer Space Telescope} observations of mid-infrared thermal emission from debris disks around early-type stars show declining dust levels with increasing system age \citep{Rieke:2005}.
The levels decline as $t_0/t$, with $t_0 \sim 150$~Myr.
Using this functional form, the solar system's interplanetary dust 3~Gyr ago was likely about 3 times brighter than it is now. 
This modest increase in brightness implies that the dust levels at that time were not high
enough to significantly modify resonant dust structures through grain-grain collisions.
Major modifications of the planets' orbital parameters, such as those described in the Nice Model, were complete by the time of the Archean Eon \citep{Gomes:2005}.
Therefore, we used the modern solar system interplanetary dust structure for the Archean solar system as well, simply scaling all dust brightnesses by a factor of 3.


\section{Background Sources \label{sec:back}}

Observations of an exoplanet system will be superimposed on a background field of spatially resolved and unresolved extragalactic and Milky Way sources that have a variety of spectra.
Strategies for minimizing confusion between point source planets and background sources need to be tested, with the goals of improving observing efficiency and guiding instrument design.
Such strategies include obtaining broadband colors or low resolution spectra, or observing the planetary system at multiple epochs to confirm that a point source is orbiting the target star.
To aid simulation and analysis of confusion mitigation strategies, we developed spectral image cubes containing extragalactic and galactic background sources separately, which can be combined with a solar system model. 
The background cubes have a larger FOV than the planetary system cubes, allowing a user to simulate the effects of proper motion by shifting the background scene.


\subsection{Extragalactic Background \label{sub:galaxies}}

Creating spectral image cubes containing extragalactic background sources at the relatively high spatial and spectral resolution of the solar system cubes -- a ``synthetic Ultra Deep Field'' (sUDF) --  was challenging.
We started from synthetic galaxy images in five bands, specifically HST/ACS F606W, F775W, and F850LP, and HST/WFC3 F125W and F160W.
They were generated by matching the luminosities and redshifts of galaxies produced in cosmological simulations to the observed population and adopting an empirical extrapolation for lower luminosities \citep{Moody:2014, Snyder:2015}. 
The sUDF FOV is \SI{12}{\arcsec} $\times$ \SI{36}{\arcsec}, about four times larger than the solar system cube.

To achieve sufficiently high spatial resolution in the sUDF, galaxies were created by applying the SUNRISE radiative transfer code \citep{Jonsson:2006} to hydro-ART galaxy formation simulations from \citet{Ceverino:2014}. These high resolution postage stamps were pieced into the larger images.
Generating the postage stamps was a computationally intensive process, and it was not feasible to create images at the very high spatial resolution of the solar system cubes. Therefore, the sUDF has larger pixels (\SI{10}{mas} $\times$ \SI{10}{mas} pixels) than the solar system cubes (\SI{3}{mas} $\times$ \SI{3}{mas} pixels).
The background galaxy fields will undersample the PSF at shorter wavelengths for large telescope diameters (see Table~\ref{tab:sampling}).
Small structures in the galaxy fields that should be spatially resolved may not be for larger telescopes/shorter wavelengths.

\begin{figure}[t!] \centering
\includegraphics[width=0.6\textwidth]{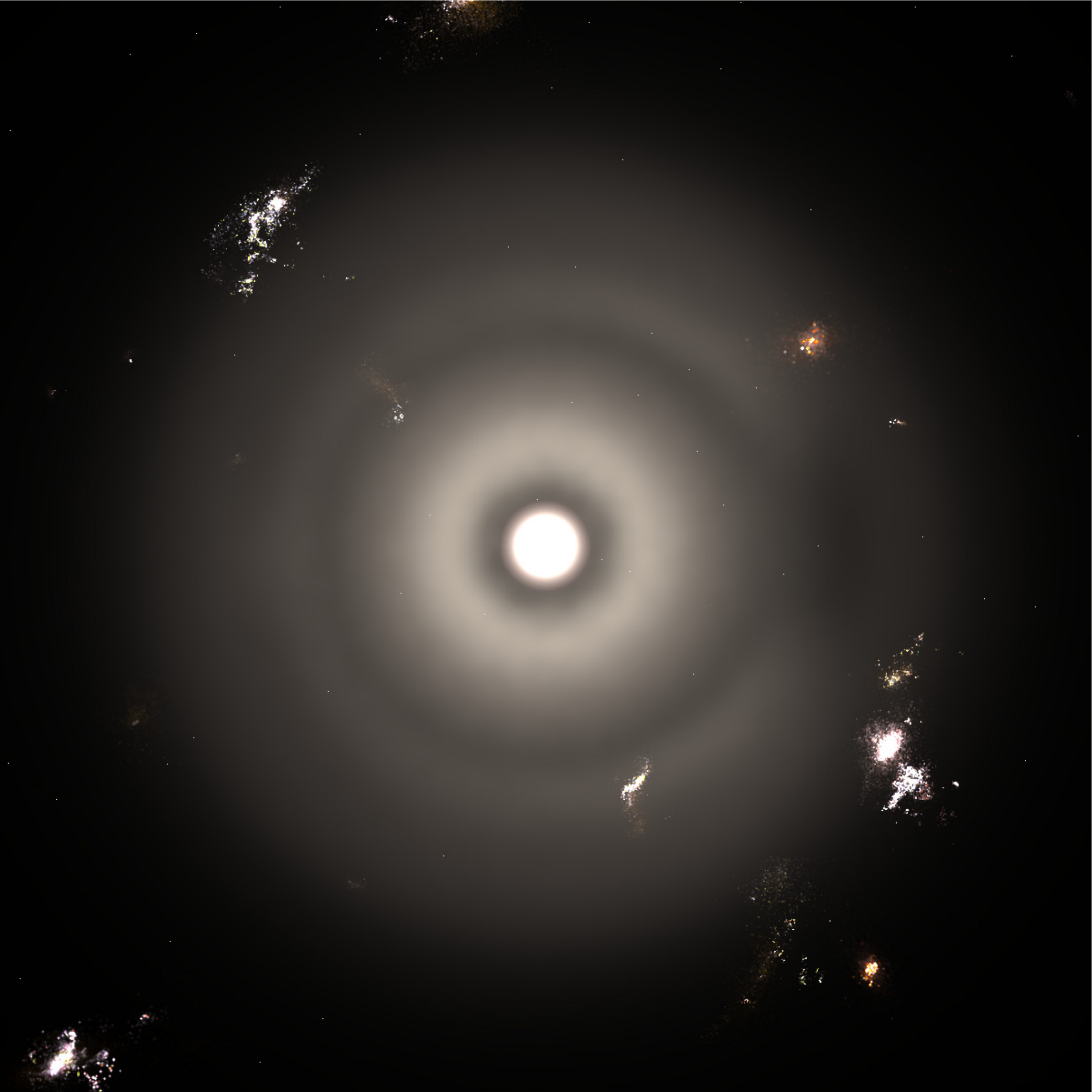}
\caption{True-color image of the modern solar system at \SI{10}{\pc} overlaid on a field of background extragalactic sources and Milky Way stars. The field of view is \SI{10}{\arcsec} $\times$ \SI{10}{\arcsec}. The planets and stars occupy single pixels and are not readily apparent without zooming in on the image.
The figure shows the solar system as it might appear situated at $b = \ang{10;;}$ and $l = \ang{0;;}$. 
Note that not all galaxies and stars included in the background field are visible at the resolution and dynamic range of this image.
}
\label{fig:galaxy_bg}
\end{figure}

We then matched the colors of each sUDF pixel to UV-Vis-IR galaxy spectra from \citet{Brown:2014}, which were shifted to redshifts between 0 and 5. 
The redshifted spectrum whose colors most closely matched the sUDF pixel colors was used to set the spectrum for that pixel, allowing us to produce $R=300$ spectral image cubes.
The Milky Way dust extinction was determined for the adopted galactic coordinates of the field ($b=10\degree$, $l=0\degree$), using the reddening maps of \citet{Schlegel:1998}. Extinctions for the five filter bands were calculated using the $R_V = 3.1$ coefficients from \citet{Schlafly:2011}. We interpolated between the bands to generate extinction coefficients at the appropriate wavelengths and applied them to the sUDF spectral image cubes.
Figure~\ref{fig:galaxy_bg} shows the modern solar system overlaid on a background field containing both galaxies and Milky Way stars.

\subsection{Milky Way Stellar Background \label{sub:mwstars}}

The density of background stars in the field of view will depend on the galactic latitude and longitude of the target star. Observations too close to the galactic plane run the risk of significant numbers of reddened background stars that form a possible source of confusion with point source planets.
The TRILEGAL (TRIdimensional modeL of thE GALaxy) stellar population synthesis code \citep{Groenewegen:2002, Girardi:2005, Girardi:2007, Vanhollebeke:2009} was used to generate a database of stars as faint as $m_{I} = 32$ for a 0.1~square-degree patch of sky at $b=10\degree$ and $l = 0\degree$.
Stellar populations from the galaxy's thin disk, thick disk, halo, and galactic bulge were included, as well as extinction by an exponential galactic dust layer.

The number of stars needed to populate the background cubes was determined from the fractional area of the background FOV (\SI{12}{\arcsec} $\times$ \SI{36}{\arcsec}) relative to the 0.1~square-degree star list.
That number of stars was randomly selected from the larger list and each placed at a random location in a stars-only field. 
A spectrum for each star was assigned based on the closest spectral match in the Pickles Atlas \citep{Pickles:1998}, scaled to the TRILEGAL brightness and interpolated onto to the Haystacks wavelength grid. For white dwarf stars, which are included in TRILEGAL, we used spectra from the Sloan Digital Sky Survey (SDSS) spectral cross-correlation templates \citep{Abazajian:2009}. These SDSS templates only cover wavelengths between about \SI{0.4}{\micron} and \SI{0.9}{\micron}, so the best-fitting blackbody curves were used to extend the templates to longer and shorter wavelengths. 

The probability of finding a background star in the stellar habitable zone increases towards smaller galactic latitudes and for galactic longitudes closer to the center of the Milky Way. Using the optimistic habitable zone boundaries for a Sun-twin star \citep[\SI{0.75}{\au} $-$ \SI{1.76}{\au};][]{Kopparapu:2013} and a face-on solar system analog viewed at a distance of \SI{10}{\pc}, we calculated the probability of finding a background star in the habitable zone.
For the worst case scenario of galactic longitude $l = 0\degree$, we find there is a $> 10 \%$ chance for observations within $b \approx 7\degree$ of the galactic plane.



\section{Simulated High-Contrast Observations\label{sec:sims}}

The Haystacks cubes can be used to produce visually appealing images showcasing the performance of high-contrast instruments, as well as high-fidelity simulations of realistic observing scenarios. 
Detailed discussions of high-contrast instrument models and post-processing techniques are deferred to a subsequent paper.
Here we provide one example in which the Haystacks modern solar system cube was used to illustrate the performance of a hypothetical Apodized Pupil Lyot Coronagraph \citep{NDiaye:2016ky} with the 12-m telescope described in the HDST report\footnote{\url{http://www.hdstvision.org/report/}}. 
The main features of this coronagraph design are an inner working angle (IWA) of 4 $\lambda$/D, an outer working angle (OWA) of 30 $\lambda$/D, throughput of \SI{18}{\percent}, and increased robustness to low-order wave front aberrations due to the relative sizes of the focal plane mask and the PSF core \citep{NDiaye:2015jz}. 

Since we were interested in creating multi-color broadband images, we integrated the fluxes of the solar system spectral image cubes over three \SI{10}{\percent} bandpasses centered around 0.4, 0.5 and 0.6~\um\ \citep[the bandpasses of the coronagraph design reported in][]{NDiaye:2015jz}. 
The solar system images were rescaled (both spatially and in flux) to 
place the star at a distance of \SI{13.5}{\pc}, so that Venus was just outside of the coronagraph's IWA and Jupiter within the OWA. At this distance, the Earth appeared at 7.2, 8.6, 10.8 $\lambda$/D for 0.5, 0.5, \SI{0.6}{\micron}, respectively. 

We began the simulation by calculating the intensity response of the coronagraph at each discrete point in the Haystacks scene. For each pixel in the scene, we introduced a tilt commensurate with off-axis pixel location in the plane of the coronagraph apodizer and then propagated the \citet{NDiaye:2016ky} design through the semi-analytical propagator presented in \citet{Soummer:2007ch}. At each position, we took the intensity of the complex electromagnetic field in the plane of the science detector, weighted it by the broadband solar system model flux in this pixel, and added this contribution to the field of unsuppressed starlight over the entire coronagraph FOV. To minimize computation time, we limited the solar system FOV to an area a slightly larger than the coronagraph field of view (e.g. 70~$\lambda$/D). 

\begin{figure}[t!] \centering
\includegraphics[width=0.5\textwidth]{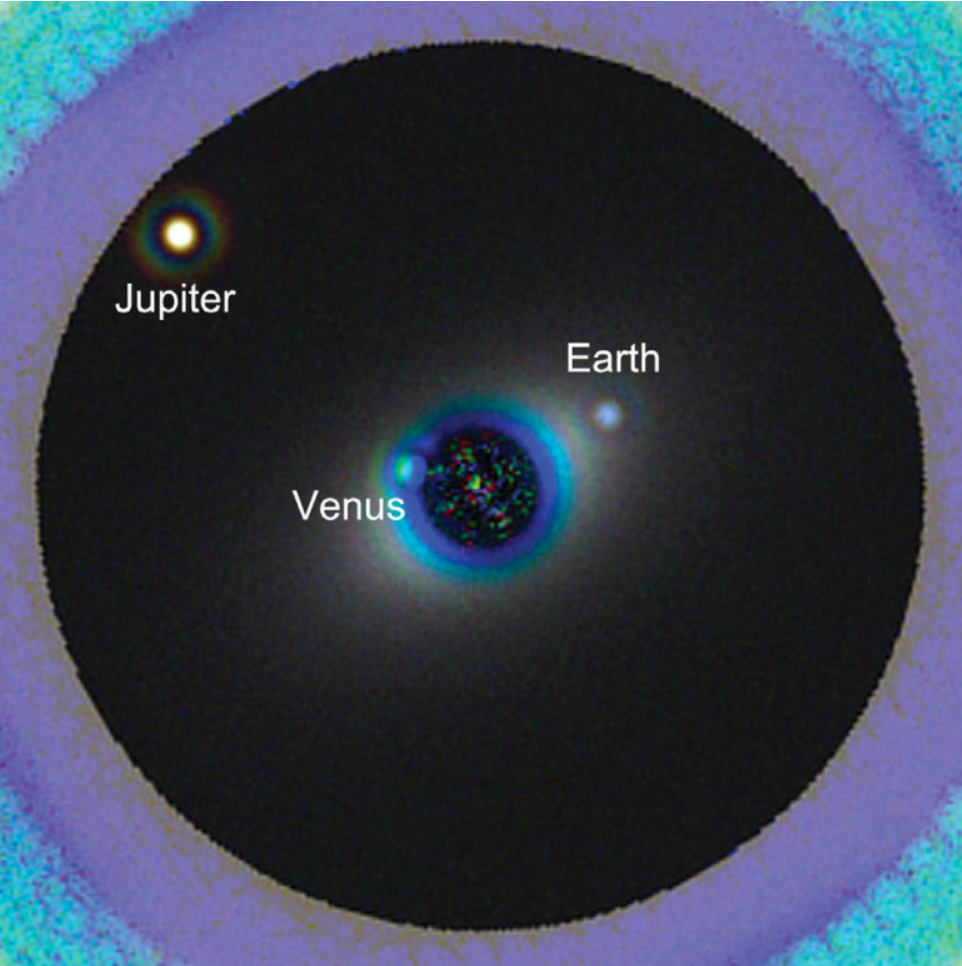}
\caption{RGB image showing the modern solar system at a distance of \SI{13.5}{\pc} imaged with an Apodized Pupil Lyot Coronagraph on the 12-m HDST.
No background galaxies or stars were included.
Ideal PSF subtraction has been assumed. The Earth with its blue color is easily detected. The color of Venus is biased towards the blue because that planet lies inside the inner working angle in the reddest exposure. Mars is undetectable. The image employs a linear brightness stretch to the outer working angle and a logarithmic stretch beyond that (where the purple-colored ring of unsuppressed starlight begins).}
\label{fig:HDST}
\end{figure}

In the resulting ``raw'' image, Earth, Venus, and Mars are hard to distinguish from the diffractive structures associated with the on-axis stellar PSF, but Jupiter is detected a high SNR. We then simulated a mock post-processing sequence by assuming that we observed a reference star of similar spectral type as the source at the center of the Haystacks scene, but devoid of circumstellar dust or exoplanets. 
For this preliminary simulation, we worked under the very optimistic assumption that there were no wavefront drifts between the target and the reference star, which would require sub-picometer stability of the whole optical system. The actual photon counts for both the target and reference stars were calculated assuming a 50\% optical throughput (including reflection losses and detector quantum efficiency) in addition to the coronagraph throughput. We further assumed an exposure time of \SI{40}{\hour}, which is similar to the time needed to characterize an exoEarth candidate with such a telescope \citep{Stark:2015er}. We subtracted the simulated reference image from the science image, resulting in the high-contrast image shown in Figure~\ref{fig:HDST}.

\section{Conclusion}

In the coming years, telescopes that can directly image a variety of exoplanetary systems with potentially habitable planets will become available. Future observatory designs are currently under development, and tools that can provide realistic analyses of how these observatories might perform are needed to constrain  design requirements. To that end, the Haystacks models were developed to provide comprehensive and realistic inputs for predicting the appearance of extrasolar planetary systems. 
We constructed realistic spectral image cubes representing the solar system at two epochs (the modern era and the Archean Eon) for use in high-fidelity simulations of high-contrast imaging and spectroscopic observations.
For the foreseeable future, the solar system will remain the best understood planetary system and therefore provides a valuable archetype for studying the properties of systems elsewhere.

The solar system models were constructed with a spatial scale of \SI{0.03}{\au} and span a wavelength range from 0.3 to \SI{2.5}{\micron}, matching the characteristics of high-contrast instruments on future large space telescopes. The fluxes from the star, the planets, and the interplanetary dust distribution are all self-consistently represented. Models at three different viewing inclinations (\SI{0}{\degree}, \SI{60}{\degree}, and \SI{90}{\degree}) are available for download. A realistic background source cube was also constructed, and can be used to simulate confusion between point-source planets and background stars and galaxies. 

A Haystack model was used to create a high-contrast imaging simulation for the HDST study. The solar system spectral image cubes could be used to simulate observations with the LUVOIR and HabEx mission concepts currently under study. By ``observing'' our solar system as an exoplanetary system, we can inform the designs of these future observatories. The scenes produced using Haystacks input models will allow us to test how challenging it will be to ``find the needles in the Haystacks'' and optimize future studies of planets across interstellar distances. 

%
%

\acknowledgements

We gratefully acknowledge helpful consultations with Ashlee Wilkins, Marc Kuchner, and Michael McElwain.
This work was supported by the NASA ROSES Origins of Solar Systems Program under proposal number 09-SSO09-0070. 
A.~R. acknowledges further support from the NASA Goddard Science Innovation Fund. G.~A. and T.~R. acknowledge support from the Virtual Planetary Laboratory, funded by the NASA Astrobiology Institute under solicitation NNH12ZDA002C and Cooperative Agreement Number NNA13AA93A. T.~R. gratefully acknowledges support from NASA through the Sagan Fellowship Program. G.~S. appreciates support from a Giacconi Fellowship at the Space Telescope Science Institute (STScI), which is operated by the Association of Universities for Research in Astronomy, Inc.\ under NASA contract NAS 5-26555. The results reported herein benefited from collaborations and/or information exchange within NASA's Nexus for Exoplanet System Science (NExSS) research coordination network, sponsored by NASA's Science Mission Directorate. 

%
%
\bibliographystyle{aasjournal}
\bibliography{haystack}

\begin{thebibliography}{}
\expandafter\ifx\csname natexlab\endcsname\relax\def\natexlab#1{#1}\fi
\providecommand{\url}[1]{\href{#1}{#1}}

\bibitem[{Abazajian {et~al.}(2009)Abazajian, Adelman-McCarthy, Ag{\"u}eros,
  Allam, Prieto, An, Anderson, Anderson, Annis, Bahcall,
  {et~al.}}]{Abazajian:2009}
Abazajian, K.~N., Adelman-McCarthy, J.~K., Ag{\"u}eros, M.~A., {et~al.} 2009,
  The Astrophysical Journal Supplement Series, 182, 543

\bibitem[{Arney {et~al.}(2016)Arney, Domagal-Goldman, Meadows, Wolf,
  Schwieterman, Charnay, Claire, H{\'e}brard, \& Trainer}]{Arney:2016}
Arney, G., Domagal-Goldman, S.~D., Meadows, V.~S., {et~al.} 2016, Astrobiology,
  16, 873

\bibitem[{{Beckwith} {et~al.}(2006){Beckwith}, {Stiavelli}, {Koekemoer},
  {Caldwell}, {Ferguson}, {Hook}, {Lucas}, {Bergeron}, {Corbin}, {Jogee},
  {Panagia}, {Robberto}, {Royle}, {Somerville}, \& {Sosey}}]{Beckwith:2006}
{Beckwith}, S.~V.~W., {Stiavelli}, M., {Koekemoer}, A.~M., {et~al.} 2006, \aj,
  132, 1729

\bibitem[{Bolcar {et~al.}(2017)Bolcar, Aloezos, Bly, Collins, Crooke, Dressing,
  Fantano, Feinberg, France, Gochar, Gong, Hylan, Jones, Linares, Postman,
  Pueyo, Roberge, Sacks, Tompkins, \& West}]{Bolcar:2017gl}
Bolcar, M.~R., Aloezos, S., Bly, V.~T., {et~al.} 2017, in UV/Optical/IR Space
  Telescopes and Instruments: Innovative Technologies and Concepts VIII
  (International Society for Optics and Photonics), 1039809

\bibitem[{Brandt \& Spiegel(2014)}]{Brandt:2014gh}
Brandt, T.~D., \& Spiegel, D.~S. 2014, in Proceedings of the National Academy
  of Sciences, 13278--13283

\bibitem[{Brown {et~al.}(2014)Brown, Moustakas, Smith, Da~Cunha, Jarrett,
  Imanishi, Armus, Brandl, \& Peek}]{Brown:2014}
Brown, M.~J., Moustakas, J., Smith, J.-D., {et~al.} 2014, The Astrophysical
  Journal Supplement Series, 212, 18

\bibitem[{{Burke} {et~al.}(2015){Burke}, {Christiansen}, {Mullally}, {Seader},
  {Huber}, {Rowe}, {Coughlin}, {Thompson}, {Catanzarite}, {Clarke}, {Morton},
  {Caldwell}, {Bryson}, {Haas}, {Batalha}, {Jenkins}, {Tenenbaum}, {Twicken},
  {Li}, {Quintana}, {Barclay}, {Henze}, {Borucki}, {Howell}, \&
  {Still}}]{Burke:2015}
{Burke}, C.~J., {Christiansen}, J.~L., {Mullally}, F., {et~al.} 2015, \apj,
  809, 8

\bibitem[{{Burrows} {et~al.}(2004){Burrows}, {Sudarsky}, \&
  {Hubeny}}]{Burrows:2004}
{Burrows}, A., {Sudarsky}, D., \& {Hubeny}, I. 2004, \apj, 609, 407

\bibitem[{{Ceverino} {et~al.}(2014){Ceverino}, {Klypin}, {Klimek},
  {Trujillo-Gomez}, {Churchill}, {Primack}, \& {Dekel}}]{Ceverino:2014}
{Ceverino}, D., {Klypin}, A., {Klimek}, E.~S., {et~al.} 2014, \mnras, 442, 1545

\bibitem[{{Cowan} {et~al.}(2011){Cowan}, {Robinson}, {Livengood}, {Deming},
  {Agol}, {A'Hearn}, {Charbonneau}, {Lisse}, {Meadows}, {Seager}, {Shields}, \&
  {Wellnitz}}]{Cowan:2011}
{Cowan}, N.~B., {Robinson}, T., {Livengood}, T.~A., {et~al.} 2011, \apj, 731,
  76

\bibitem[{{Cox}(2000)}]{Cox:2000}
{Cox}, A.~N. 2000, {Allen's Astrophysical Quantities, 4th Ed.} (AIP Press;
  Springer)

\bibitem[{{Crisp}(1997)}]{Crisp:1997}
{Crisp}, D. 1997, \grl, 24, 571

\bibitem[{{Des Marais} {et~al.}(2002){Des Marais}, {Harwit}, {Jucks},
  {Kasting}, {Lin}, {Lunine}, {Schneider}, {Seager}, {Traub}, \&
  {Woolf}}]{DesMarais:2002}
{Des Marais}, D.~J., {Harwit}, M.~O., {Jucks}, K.~W., {et~al.} 2002,
  Astrobiology, 2, 153

\bibitem[{{Dohnanyi}(1969)}]{Dohnanyi:1969}
{Dohnanyi}, J.~S. 1969, \jgr, 74, 2531

\bibitem[{{Domagal-Goldman} {et~al.}(2008){Domagal-Goldman}, {Kasting},
  {Johnston}, \& {Farquhar}}]{DomagalGoldman:2008}
{Domagal-Goldman}, S.~D., {Kasting}, J.~F., {Johnston}, D.~T., \& {Farquhar},
  J. 2008, Earth and Planetary Science Letters, 269, 29

\bibitem[{{Donahue} \& {Hodges}(1992)}]{Donahue:1992}
{Donahue}, T.~M., \& {Hodges}, Jr., R.~R. 1992, \jgr, 97, 6083

\bibitem[{Driese {et~al.}(2011)Driese, Jirsa, Ren, Brantley, Sheldon, Parker,
  \& Schmitz}]{Driese:2011}
Driese, S.~G., Jirsa, M.~A., Ren, M., {et~al.} 2011, Precambrian Research, 189,
  1

\bibitem[{{Fortney} {et~al.}(2010){Fortney}, {Baraffe}, \&
  {Militzer}}]{Fortney:2010}
{Fortney}, J.~J., {Baraffe}, I., \& {Militzer}, B. 2010, {Giant Planet Interior
  Structure and Thermal Evolution}, ed. S.~{Seager} (University of Arizona
  Press), 397--418

\bibitem[{{Girardi} {et~al.}(2005){Girardi}, {Groenewegen}, {Hatziminaoglou},
  \& {da Costa}}]{Girardi:2005}
{Girardi}, L., {Groenewegen}, M.~A.~T., {Hatziminaoglou}, E., \& {da Costa}, L.
  2005, \aap, 436, 895

\bibitem[{{Girardi} \& {Marigo}(2007)}]{Girardi:2007}
{Girardi}, L., \& {Marigo}, P. 2007, in Astronomical Society of the Pacific
  Conference Series, Vol. 378, Why Galaxies Care About AGB Stars: Their
  Importance as Actors and Probes, ed. F.~{Kerschbaum}, C.~{Charbonnel}, \&
  R.~F. {Wing}, 20

\bibitem[{{Gomes} {et~al.}(2005){Gomes}, {Levison}, {Tsiganis}, \&
  {Morbidelli}}]{Gomes:2005}
{Gomes}, R., {Levison}, H.~F., {Tsiganis}, K., \& {Morbidelli}, A. 2005, \nat,
  435, 466

\bibitem[{{Groenewegen} {et~al.}(2002){Groenewegen}, {Girardi},
  {Hatziminaoglou}, {Benoist}, {Olsen}, {da Costa}, {Arnouts}, {Madejsky},
  {Mignani}, {Rit{\'e}}, {Sikkema}, {Slijkhuis}, \&
  {Vandame}}]{Groenewegen:2002}
{Groenewegen}, M.~A.~T., {Girardi}, L., {Hatziminaoglou}, E., {et~al.} 2002,
  \aap, 392, 741

\bibitem[{Hamano {et~al.}(2013)Hamano, Abe, \& Genda}]{Hamano:2013}
Hamano, K., Abe, Y., \& Genda, H. 2013, Nature, 497, 607

\bibitem[{Hedman \& Stark(2015)}]{Hedman:2015ie}
Hedman, M.~M., \& Stark, C.~C. 2015, The Astrophysical Journal, 811, 67

\bibitem[{{Hinz} {et~al.}(2016){Hinz}, {Defr{\'e}re}, {Skemer}, {Bailey},
  {Stone}, {Spalding}, {Vaz}, {Pinna}, {Puglisi}, {Esposito}, {Montoya},
  {Downey}, {Leisenring}, {Durney}, {Hoffmann}, {Hill}, {Millan-Gabet},
  {Mennesson}, {Danchi}, {Morzinski}, {Grenz}, {Skrutskie}, \&
  {Ertel}}]{Hinz:2016}
{Hinz}, P.~M., {Defr{\'e}re}, D., {Skemer}, A., {et~al.} 2016, in Optical and
  Infrared Interferometry and Imaging V, Vol. 9907, 990704

\bibitem[{{Izon} {et~al.}(2015){Izon}, {Zerkle}, {Zhelezinskaia}, {Farquhar},
  {Newton}, {Poulton}, {Eigenbrode}, \& {Claire}}]{Izon:2015}
{Izon}, G., {Zerkle}, A.~L., {Zhelezinskaia}, I., {et~al.} 2015, Earth and
  Planetary Science Letters, 431, 264

\bibitem[{{Jonsson}(2006)}]{Jonsson:2006}
{Jonsson}, P. 2006, \mnras, 372, 2

\bibitem[{{Kaltenegger} \& {Traub}(2009)}]{Kaltenegger:2009}
{Kaltenegger}, L., \& {Traub}, W.~A. 2009, \apj, 698, 519

\bibitem[{{Karkoschka}(1998)}]{Karkoschka:1998}
{Karkoschka}, E. 1998, \icarus, 133, 134

\bibitem[{{Kelsall} {et~al.}(1998){Kelsall}, {Weiland}, {Franz}, {Reach},
  {Arendt}, {Dwek}, {Freudenreich}, {Hauser}, {Moseley}, {Odegard},
  {Silverberg}, \& {Wright}}]{Kelsall:1998}
{Kelsall}, T., {Weiland}, J.~L., {Franz}, B.~A., {et~al.} 1998, \apj, 508, 44

\bibitem[{{Kopparapu} {et~al.}(2013){Kopparapu}, {Ramirez}, {Kasting}, {Eymet},
  {Robinson}, {Mahadevan}, {Terrien}, {Domagal-Goldman}, {Meadows}, \&
  {Deshpande}}]{Kopparapu:2013}
{Kopparapu}, R.~K., {Ramirez}, R., {Kasting}, J.~F., {et~al.} 2013, \apj, 765,
  131

\bibitem[{{Kuchner}(2012)}]{Kuchner:2012}
{Kuchner}, M. 2012, {ZODIPIC: Zodiacal Cloud Image Synthesis},  Astrophysics
  Source Code Library, ascl:1202.002

\bibitem[{Kuchner \& Holman(2003)}]{Kuchner:2003ja}
Kuchner, M.~J., \& Holman, M.~J. 2003, The Astrophysical Journal, 588, 1110

\bibitem[{{Kuchner} \& {Stark}(2010)}]{KuchnerStark:2010}
{Kuchner}, M.~J., \& {Stark}, C.~C. 2010, \aj, 140, 1007

\bibitem[{{Meadows} \& {Crisp}(1996)}]{Meadows:1996}
{Meadows}, V.~S., \& {Crisp}, D. 1996, \jgr, 101, 4595

\bibitem[{Mennesson {et~al.}(2016)Mennesson, Gaudi, Seager, Cahoy,
  Domagal-Goldman, Feinberg, Guyon, Kasdin, Marois, Mawet, Tamura, Mouillet,
  Prusti, Quirrenbach, Robinson, Rogers, Scowen, Somerville, Stapelfeldt,
  Stern, Still, Turnbull, Booth, Kiessling, Kuan, \&
  Warfield}]{Mennesson:2016fp}
Mennesson, B., Gaudi, S., Seager, S., {et~al.} 2016, in Proceedings of the
  SPIE, Jet Propulsion Lab. (United States) (International Society for Optics
  and Photonics), 99040L

\bibitem[{Moody {et~al.}(2014)Moody, Guo, Mandelker, Ceverino, Mozena, Koo,
  Dekel, \& Primack}]{Moody:2014}
Moody, C.~E., Guo, Y., Mandelker, N., {et~al.} 2014, Monthly Notices of the
  Royal Astronomical Society, 444, 1389

\bibitem[{N'Diaye {et~al.}(2015)N'Diaye, Pueyo, \& Soummer}]{NDiaye:2015jz}
N'Diaye, M., Pueyo, L., \& Soummer, R. 2015, \apj, 799, 225

\bibitem[{N'Diaye {et~al.}(2016)N'Diaye, Soummer, Pueyo, Carlotti, Stark, \&
  Perrin}]{NDiaye:2016ky}
N'Diaye, M., Soummer, R., Pueyo, L., {et~al.} 2016, \apj, 818, 163

\bibitem[{{Noecker} \& {Kuchner}(2010)}]{Noecker:2010}
{Noecker}, C., \& {Kuchner}, M. 2010, ArXiv e-prints, arXiv:1012.0977

\bibitem[{Pickles(1998)}]{Pickles:1998}
Pickles, A. 1998, Publications of the Astronomical Society of the Pacific, 110,
  863

\bibitem[{{Rayner} {et~al.}(2009){Rayner}, {Cushing}, \& {Vacca}}]{Rayner:2009}
{Rayner}, J.~T., {Cushing}, M.~C., \& {Vacca}, W.~D. 2009, \apjs, 185, 289

\bibitem[{{Rieke} {et~al.}(2005){Rieke}, {Su}, {Stansberry}, {Trilling},
  {Bryden}, {Muzerolle}, {White}, {Gorlova}, {Young}, {Beichman},
  {Stapelfeldt}, \& {Hines}}]{Rieke:2005}
{Rieke}, G.~H., {Su}, K.~Y.~L., {Stansberry}, J.~A., {et~al.} 2005, \apj, 620,
  1010

\bibitem[{{Roberge} {et~al.}(2012){Roberge}, {Chen}, {Millan-Gabet},
  {Weinberger}, {Hinz}, {Stapelfeldt}, {Absil}, {Kuchner}, \&
  {Bryden}}]{Roberge:2012}
{Roberge}, A., {Chen}, C.~H., {Millan-Gabet}, R., {et~al.} 2012, \pasp, 124,
  799

\bibitem[{{Robinson} {et~al.}(2011){Robinson}, {Meadows}, {Crisp}, {Deming},
  {A'Hearn}, {Charbonneau}, {Livengood}, {Seager}, {Barry}, {Hearty},
  {Hewagama}, {Lisse}, {McFadden}, \& {Wellnitz}}]{Robinson:2011}
{Robinson}, T.~D., {Meadows}, V.~S., {Crisp}, D., {et~al.} 2011, Astrobiology,
  11, 393

\bibitem[{Savransky {et~al.}(2010)Savransky, Kasdin, \&
  Cady}]{Savransky:2010fd}
Savransky, D., Kasdin, N.~J., \& Cady, E. 2010, Publications of the
  Astronomical Society of Pacific, 122, 401

\bibitem[{{Savransky} {et~al.}(2009){Savransky}, {Kasdin}, \&
  {Vanderbei}}]{Savransky:2009}
{Savransky}, D., {Kasdin}, N.~J., \& {Vanderbei}, R.~J. 2009, in \procspie,
  Vol. 7440, Techniques and Instrumentation for Detection of Exoplanets IV,
  744015

\bibitem[{Schlafly \& Finkbeiner(2011)}]{Schlafly:2011}
Schlafly, E.~F., \& Finkbeiner, D.~P. 2011, The Astrophysical Journal, 737, 103

\bibitem[{Schlegel {et~al.}(1998)Schlegel, Finkbeiner, \&
  Davis}]{Schlegel:1998}
Schlegel, D.~J., Finkbeiner, D.~P., \& Davis, M. 1998, The Astrophysical
  Journal, 500, 525

\bibitem[{Schneider {et~al.}(2009)Schneider, Weinberger, Becklin, Debes, \&
  Smith}]{Schneider:2009fk}
Schneider, G., Weinberger, A.~J., Becklin, E.~E., Debes, J.~H., \& Smith, B.~A.
  2009, The Astronomical Journal, 137, 53

\bibitem[{{Smith} {et~al.}(2014){Smith}, {Claire}, {Catling}, \&
  {Zahnle}}]{Smith:2014}
{Smith}, M.~L., {Claire}, M.~W., {Catling}, D.~C., \& {Zahnle}, K.~J. 2014,
  \icarus, 231, 51

\bibitem[{Snyder {et~al.}(2015)Snyder, Torrey, Lotz, Genel, McBride,
  Vogelsberger, Pillepich, Nelson, Sales, Sijacki, {et~al.}}]{Snyder:2015}
Snyder, G.~F., Torrey, P., Lotz, J.~M., {et~al.} 2015, Monthly Notices of the
  Royal Astronomical Society, 454, 1886

\bibitem[{Soummer {et~al.}(2007)Soummer, Pueyo, Sivaramakrishnan, \&
  Vanderbei}]{Soummer:2007ch}
Soummer, R., Pueyo, L., Sivaramakrishnan, A., \& Vanderbei, R.~J. 2007, Optics
  Express, 15, 15935

\bibitem[{{Stark}(2011)}]{Stark:2011}
{Stark}, C.~C. 2011, \aj, 142, 123

\bibitem[{{Stark} \& {Kuchner}(2009)}]{StarkKuchner:2009}
{Stark}, C.~C., \& {Kuchner}, M.~J. 2009, \apj, 707, 543

\bibitem[{Stark {et~al.}(2015)Stark, Roberge, Mandell, Clampin,
  Domagal-Goldman, McElwain, \& Stapelfeldt}]{Stark:2015er}
Stark, C.~C., Roberge, A., Mandell, A., {et~al.} 2015, \apj, 808, 149

\bibitem[{Stark {et~al.}(2014)Stark, Roberge, Mandell, \&
  Robinson}]{Stark:2014dt}
Stark, C.~C., Roberge, A., Mandell, A., \& Robinson, T.~D. 2014, The
  Astrophysical Journal, 795, 122

\bibitem[{{Trainer} {et~al.}(2006){Trainer}, {Pavlov}, {Dewitt}, {Jimenez},
  {McKay}, {Toon}, \& {Tolbert}}]{Trainer:2006}
{Trainer}, M.~G., {Pavlov}, A.~A., {Dewitt}, H.~L., {et~al.} 2006, Proceedings
  of the National Academy of Science, 103, 18035

\bibitem[{{Turnbull} {et~al.}(2012){Turnbull}, {Glassman}, {Roberge}, {Cash},
  {Noecker}, {Lo}, {Mason}, {Oakley}, \& {Bally}}]{Turnbull:2012}
{Turnbull}, M.~C., {Glassman}, T., {Roberge}, A., {et~al.} 2012, \pasp, 124,
  418

\bibitem[{{Vanhollebeke} {et~al.}(2009){Vanhollebeke}, {Groenewegen}, \&
  {Girardi}}]{Vanhollebeke:2009}
{Vanhollebeke}, E., {Groenewegen}, M.~A.~T., \& {Girardi}, L. 2009, \aap, 498,
  95

\bibitem[{Way {et~al.}(2016)Way, Del~Genio, Kiang, Sohl, Grinspoon, Aleinov,
  Kelley, \& Clune}]{Way:2016}
Way, M.~J., Del~Genio, A.~D., Kiang, N.~Y., {et~al.} 2016, Geophysical Research
  Letters, 43, 8376

\bibitem[{Zerkle {et~al.}(2012)Zerkle, Claire, Domagal-Goldman, Farquhar, \&
  Poulton}]{Zerkle:2012}
Zerkle, A.~L., Claire, M.~W., Domagal-Goldman, S.~D., Farquhar, J., \& Poulton,
  S.~W. 2012, Nature Geoscience, 5, 359

\end{thebibliography}

%
%

%
%

\end{document}